\documentclass[pra,aps,twocolumn,floatfix,showpacs,superscriptaddress]{revtex4-1}

\usepackage{graphicx}                             
\graphicspath{ {./_Figures/} }

\usepackage{epsfig}
\usepackage{epstopdf}
\usepackage{color}
\definecolor{red}{rgb}{1.00,0.00,0.00}
\usepackage{changes}

\usepackage[breaklinks]{hyperref}
\usepackage[all]{hypcap}

\hypersetup{
	plainpages=false,
	unicode=false,          
	pdftoolbar=true,        
	pdfmenubar=true,        
	pdffitwindow=false,     
	pdfstartview={FitH},    
	pdftitle={Phys. Rev. B},    
	pdfauthor={Thorsten Hesjedal},     
	pdfproducer={University of Oxford}, 
	pdfnewwindow=true,      
	linktoc=section,
	colorlinks=true,       
	linkcolor=blue,          
	citecolor=blue,        
	filecolor=magenta,      
	urlcolor=blue           
}

\date{\today}

\begin{document}

\title{Microscopic effects of Dy-doping in the topological insulator Bi$_2$Te$_3$}

\author{L.~B. Duffy}
\affiliation{Department of Physics, Clarendon Laboratory, University of Oxford, Oxford, OX1~3PU, United Kingdom}
\affiliation{ISIS, Rutherford Appleton Laboratory, Harwell Science and Innovation Campus, Science and Technology Facilities Council, Oxon OX11~0QX, United Kingdom}

\author{N.-J. Steinke}
\email[Corresponding author: ]{nina-juliane.steinke@stfc.ac.uk}
\affiliation{ISIS, Rutherford Appleton Laboratory, Harwell Science and Innovation Campus, Science and Technology Facilities Council, Oxon OX11~0QX, United Kingdom}

\author{J. A. Krieger}
\affiliation{Laboratory for Muon Spin Spectroscopy, Paul Scherrer Institut, Villigen, Switzerland}
\affiliation{Laboratorium f\"ur Festk\"orperphysik, ETH-H\"onggerberg, CH-8093 Z\"urich, Switzerland}

\author{A.~I. Figueroa}
\affiliation{Magnetic Spectroscopy Group, Diamond Light Source, Didcot, OX11~0DE, United Kingdom}

\author{K. Kummer}
\affiliation{European Synchrotron Radiation Facility, BP~220, 38043 Grenoble Cedex, France}

\author{T. Lancaster}
\affiliation{Centre for Materials Physics, Durham University, Durham, DH1~3LE, United Kingdom}

\author{S.~R. Giblin}
\affiliation{School of Physics and Astronomy, Cardiff University, Queen's Buildings, The Parade, Cardiff, CF24~3AA, United Kingdom}

\author{F.~L. Pratt}
\affiliation{ISIS, Rutherford Appleton Laboratory, Harwell Science and Innovation Campus, Science and Technology Facilities Council, Oxon OX11~0QX, United Kingdom}

\author{S.~J. Blundell}
\affiliation{Department of Physics, Clarendon Laboratory, University of Oxford, Oxford, OX1~3PU, United Kingdom}

\author{T. Prokscha}
\affiliation{Laboratory for Muon Spin Spectroscopy, Paul Scherrer Institut, Villigen, Switzerland}

\author{A. Suter}
\affiliation{Laboratory for Muon Spin Spectroscopy, Paul Scherrer Institut, Villigen, Switzerland}

\author{S. Langridge}
\affiliation{ISIS, Rutherford Appleton Laboratory, Harwell Science and Innovation Campus, Science and Technology Facilities Council, Oxon OX11~0QX, United Kingdom}


\author{V. N. Strocov} \affiliation{Swiss Light Source, Paul Scherrer Institute, CH-5232 Villigen PSI, Switzerland}

\author{Z. Salman}
\affiliation{Laboratory for Muon Spin Spectroscopy, Paul Scherrer Institut, Villigen, Switzerland}

\author{G. \surname{van der Laan}}
\affiliation{Magnetic Spectroscopy Group, Diamond Light Source, Didcot, OX11~0DE, United Kingdom}

\author{T. Hesjedal}
\affiliation{Department of Physics, Clarendon Laboratory, University of Oxford, Oxford, OX1~3PU, United Kingdom}

\begin{abstract}
Magnetic doping with transition metal ions is the most widely used approach to break time-reversal symmetry in a topological insulator (TI) --- a prerequisite for unlocking the TI's exotic potential.
Recently, we reported the doping of Bi$_2$Te$_3$ thin films with rare earth ions, which, owing to their large magnetic moments, promise commensurately large magnetic gap openings in the topological surface states (TSS). 
However, only when doping with Dy has a sizable gap been observed in angle-resolved photoemission spectroscopy, which persists up to room-temperature.
Although disorder alone could be ruled out as a cause of the topological phase transition, a fundamental understanding of the magnetic and electronic properties of Dy-doped Bi$_2$Te$_3$ remained elusive.
Here, we present an X-ray magnetic circular dichroism, polarized neutron reflectometry, muon spin rotation, and resonant photoemission (ResPE) study of the microscopic magnetic and electronic properties.
We find that the films are not simply paramagnetic but that instead the observed behavior can be well explained by the assumption of slowly fluctuating, inhomogeneous, magnetic patches with increasing volume fraction as the temperature decreases. At liquid helium temperatures, a large effective magnetization can be easily introduced by the application of moderate magnetic fields, implying that this material is very suitable for proximity coupling to an underlying ferromagnetic insulator or in a heterostructure with transition metal-doped layers. However, the introduction of some charge carriers by the Dy-dopants cannot be excluded at least in these highly doped samples. Nevertheless, we find that the magnetic order is not mediated via the conduction channel in these rare earth doped samples and therefore magnetic order and carrier concentration are expected to be independently controllable. This is not generally the case for transition metal doped topological insulators. 
\end{abstract}

\pacs{75.30.Hx; 78.70.Dm; 75.50.Pp; 73.61.Ng}


\date{\today}
\maketitle







\section{\label{sec:intro}Introduction}

Three-dimensional (3D) topological insulators (TIs) \cite{Kane2005,Fu2007,Qi2008,Zhang2009} have captured the attention of the condensed matter physics community owing to the intriguing properties of their spin-momentum locked topological surface states (TSSs).
The TSS is protected by time-reversal symmetry (TRS), making it robust against backscattering from non-magnetic impurities \cite{Roushan2009}.
To observe exotic quantum phenomena such as the quantum anomalous Hall effect (QAHE) \cite{Yu2010, Liu2008, Wang2014, Wang2013_CrBiSbTe_Theory_QAHE, Liu2013_InPlaneQAHE},
TRS has to be broken and a controlled gap has to be introduced in the linearly dispersing TSS Dirac cone.
This is achieved by doping with magnetic impurities \cite{He2014_QAHE}, which means that the QAHE can be observed in the absence of an applied magnetic field.
Long-range ferromagnetically ordered TIs exhibit a finite gap of tens of meV at the Dirac point \cite{Yu2010}, transforming the massless Dirac Fermion state to a massive one \cite{Chen2010,Sengupta2014_influence}. 

Ferromagnetic doping in the common 3D TIs Bi$_2$Se$_3$ and Bi$_2$Te$_3$ is focused on the $3d$ transition metal ions such as Cr \cite{Haazen2012,Collins2014_MagneticCrBS} or Mn \cite{Chen2010,Zhang2012a,Watson2013} or V \cite{Chang2013, Chang2015}.
In Cr- and V-doped (Bi,Sb)$_2$Te$_3$ thin films, the QAHE has been observed at mK temperatures \cite{Chang2013,Chang2015}, despite a magnetic ordering temperature in those samples of 15~K in case of Cr doping \cite{Chang2013} and about twice the value for V doping \cite{Chang2015}.
This highlights an important problem for the QAHE: The deterioration of the electric transport properties through the presence of magnetic doping, often attributed to the presence of non-topological impurity bands near the Fermi-surface which allow bulk conduction \cite{Peixoto2016,Krieger2017}. In consequence, very low temperatures are necessary, both to enhance magnetic order and to suppress the dissipative transport channels \cite{Kou2014}.
In order to minimize the adverse effects of the magnetic impurities, low doping concentrations are desirable. As the size of the Dirac gap increases with the magnetic moment \cite{Chen2010}, 
the doping concentration can be reduced by using a larger magnetic moment.

In our recent work, we have shown that $4f$ rare earth (RE) ions, such as the high moment ions Ho \cite{Harrison2015_Ho} and Dy \cite{Harrison2014_Gd,Harrison2015_Gd,Harrison_SciRep} with effective magnetic moments ($m_\mathrm{eff}$) of up to $\sim$10.65~$\mu_\mathrm{B}$, can be used to dope thin films of Bi$_2$Te$_3$ up to $\sim$35\% (in \% of the Bi sites for substitutional doping) without loss of crystalline quality.
Furthermore, as shown for Gd-doped Bi$_2$Te$_3$ thin films \cite{Harrison2014_Gd}, the isoelectronic substitutional doping of RE$^{3+}$ ions on Bi sites should not introduce any additional charge carriers.

Owing to their well-shielded $4f$ shells, RE ions behave like isolated magnetic moments and the materials thus remain paramagnetic \cite{Jensen1991}.
Despite this absence of long-range ferromagnetic order in Dy-doped Bi$_2$Te$_3$, we have observed a gapped TSS band in angle-resolved photoemission spectroscopy (ARPES) up to room-temperature \cite{Harrison_SciRep}, 
which is unexpected but not unprecedented \cite{Xu2012,Franz2013_TIBook}.
In fact, in Cr:Bi$_{2}$Se$_3$ thin films, a gapped TSS has also been observed in the absence of long-range ferromagnetic order, which the authors attributed to the inhomogeneous doping and the formation of Cr clusters which lead to short-range ferromagnetic order \cite{2014_ARPESGap_CrClusters}.
This illustrates the important role disorder and subsequent impurity scattering appears to play in magnetically doped TIs in general \cite{Alpichshev2012_STM,Black2012_strong}.

The experimentally determined gap sizes in Mn-doped Bi$_2$Se$_3$ and Bi$_2$Te$_3$ are well above the values predicted by theory, and the gap opening persist to temperatures well above the magnetic transition temperatures \cite{Henk2012b,Xu2012}, casting doubts on their magnetic origins.
However, in RE-doped Bi$_2$Te$_3$ thin films, we only found a gap opening in the case of Dy doping \cite{Harrison_SciRep}, but not for Ho- or Gd-doping, despite the very similar structural properties (at the same doping levels) \cite{Li2013,Harrison2014_Gd,Harrison2014_Cleave}. It appears that the microscopic nature of Dy doping is more complex, as evidenced by its magnetic properties as well \cite{Harrison2015_Gd}.

Here we report a microscopic study of the magnetic properties of Dy-doped Bi$_2$Te$_3$ thin films using x-ray circular magnetic dichroism (XMCD), polarized neutron reflectometry (PNR), and muon-spin rotation ($\mu^+$SR), and of the effect of Dy-doping on the electronic band structure using soft x-ray ARPES.
Temperature and magnetic-field dependent XMCD at the $M_{4,5}$ edges of Dy and Te give insight into the magnetic ordering and coupling in the films, whereas PNR allows for a precise determination of the doping concentration and the effective magnetization of the material as a function of depth \cite{Marrows2009}. 
$\mu^+$SR is a well-established technique to study local magnetic order, which has been very successful for understanding dilute magnetic semiconductors \cite{Saadaoui2016,Dunsiger2010}. 
Finally, we used soft x-ray ARPES to gain chemical selectivity for the study of the electronic characteristics of Dy dopants in Bi$_2$Te$_3$.

We find that Dy-doped Bi$_2$Te$_3$, while not ferromagnetic itself, offers a large, induced magnetic polarization at small external fields with slow fluctuations of the very large moments. We see some evidence for dopant induced excitations close to the Fermi energy but no evidence that these contribute to the (short range) magnetic ordering. These properties make Dy-doped TIs interesting materials for use as functional layers in complex TI heterostructures.  

\section{\label{sec:exp}Experimental Methods}
\textbf{Sample growth and previous characterization:}
(Dy$_x$Bi$_{1-x}$)$_2$Te$_3$ thin films, with $x$ denoting the substitutional Dy concentration (as a fraction of the Bi sites), were grown by molecular beam epitaxy (MBE) on Al$_2$O$_3$(0001) ($c$-plane sapphire) substrates \cite{Harrison2015_Gd}. The rhombohedral films were prepared using a two-step growth recipe with substrate temperatures of 250$^\circ$C and 300$^\circ$C. The films are free of secondary phases as confirmed by x-ray diffraction. 
Using \textit{in situ} reflection high-energy electron diffraction (RHEED) we confirmed that streak-like diffraction patterns remained observable during growth.
The growth recipe and the structural properties of the films are described in detail in Refs.\ \cite{Harrison2015_Gd,Harrison_SciRep}. The films investigated there had a Dy concentration of $x=0.055$ and 0.113.

For the lower Dy concentration, the TSS bandstructure remains intact and shows a linear dispersion, whereas for the higher concentration a sizable gap was observed in ARPES \cite{Harrison_SciRep}, suggesting a possible change in magnetic properties.

\textbf{X-ray spectroscopy:}
X-ray absorption spectroscopy (XAS) and x-ray magnetic circular dichroism (XMCD) measurements were carried out at beamline ID32 \cite{Kummer2014} at the European Synchrotron Radiation Facility (ESRF), Grenoble (France). All measurements were performed on an \emph{in-situ} cleaved $(\mathrm{Dy}_{0.31}\mathrm{Bi}_{0.69})_2\mathrm{Te}_3$ sample (the same sample as used for the neutron measurements) in total-electron-yield (TEY) mode, giving a sampling depth of the order of 3-5~nm.
We studied the Te $M_{4,5}$ and the Dy $M_{4,5}$ edges as a function of temperature and in-plane applied magnetic field.

\textbf{Polarized neutron reflectometry:}
The structural and magnetic depth profiles of two samples with high Dy concentration were explored using polarized neutron reflectometry (PNR) without polarization analysis at the Polref beamline at the ISIS neutron and muon source (Rutherford Appleton Laboratory, UK). Measurements were carried out at room temperature and low temperatures (3~K and 5~K) and low and high magnetic fields (0.02~T and 0.7~T), applied in the sample plane, parallel to the neutron spin polarization. Reflectivity curves were acquired for the two spin eigenstates (commonly referred to as spin-up and spin-down) of the incoming beam. In this configuration the reflectivity measurement is sensitive to the structural depth profile as well as the magnetic component parallel to the applied field. In particular the neutron senses the total magnetic induction $\mathbf{B(z)}$ as a function of depth within the sample.
Model fitting was carried out using the GenX package \cite{Bjorck2007} based on an optical transfer matrix approach \cite{Blundell1992} and a differential evolution fitting algorithm. 

\textbf{Muon spin spectroscopy:}
Transverse and longitudinal field muon-spin rotation ($\mu^{+}$SR) measurements were carried out using the low-energy muon (LEM) beamline at S$\mu$S \cite{prokscha}. 
The LEM beamline produces very slow, spin-polarized, positive muons with tunable energies between 0 and 30~keV \cite{morenzoni,morenzoni2}. Muon spin spectroscopy is a very sensitive probe of the local (but not element specific) magnetic order. Dynamics can be probed on the $\mu s$ timescale.
For the measurements, the samples were glued to a Ni plated Al sample holder with a small magnetic field applied perpendicular to the surface of the sample. Measurements were made on two Dy:Bi$_2$Te$_3$ films of different Dy concentrations, one of which is equivalent to the one used for XMCD and PNR. All data analysis was carried out using the WiMDA program \cite{WiMDA}.

\textbf{Soft x-ray angle resolved photoemission spectroscopy:}
Soft x-ray ARPES measurements were performed at the ADRESS beamline of the Swiss Light Source at the Paul Scherrer Institut, Switzerland~\cite{Strocov2010}.
The experiment was carried out with linear horizontal polarized light at a pressure below 10$^{-10}$~mbar and temperatures below 11.5~K on the same sample that was investigated in the $\mu^+$SR experiment. The sample was \emph{in-situ} cleaved before the measurements.
The combined analyzer and beamline resolution at photon energies around 1.3~keV was better than 170 meV. 
The ARPES setup is described elsewhere~\cite{Strocov2014}. Supporting XAS measurements were conducted by measuring the TEY.

\section{\label{sec:res}Results}

\subsection{\label{sec:disStruct}Structural properties}

The structural properties of the samples investigated here are commensurate with those reported in our previous investigations \cite{Harrison2014_Gd, Harrison2015_Gd}: X-ray diffraction measurements show single phase diffraction patterns with no contamination phases even at the highest doping concentration of $x = 0.35$. The $c$-axis lattice parameter shifts slightly from $30.4$~{\AA} in an undoped Bi$_2$Te$_3$ thin film grown in the same chamber to 30.7~{\AA} for the main sample under investigation ($x = 0.31$). As the ionic radius of Dy$^{3+}$ (105~pm) is similar, but slightly smaller than that of the host cation Bi$^{3+}$ (117~pm), a fraction of the Dy does not dope substitutionally, but is instead likely to accumulate in the van der Waals gap.

Neutron reflectivity measurements are sensitive to the replacement of Bi ions by Dy as the bound coherent nuclear scattering length, $\mathtt{b}_c$, is much stronger for Dy ($\mathtt{b}_c^{\mathtt{Dy}} = 16.9$~fm) than for Bi ($\mathtt{b}_c^{\mathtt{Bi}} = 8.5$~fm).
We are able to estimate the doping concentration from the fits of the high temperature polarized neutron data to $(\mathrm{Dy}_{0.31}\mathrm{Bi}_{0.69})_2\mathrm{Te}_3$ for the main sample under investigation and to $(\mathrm{Dy}_{0.35}\mathrm{Bi}_{0.65})_2\mathrm{Te}_3$ for the more highly doped sample, i.e., $x=0.31$ and $x=0.35$, respectively.
In our estimate of the Dy concentration we make use of the assumption that the accumulation in the van der Waals gap does not significantly alter the number of atoms present per unit volume (the increased number of atoms in the gap is offset mostly by the expansion of the lattice). As previously reported \cite{Harrison2014_Gd,Harrison2015_Gd}, the crystal structure is stable even when incorporating high concentrations of Dy with both substitution and accumulation in the van der Waals gap, but without the formation of parasitic phases.

\subsection{\label{sec:disMag}Magnetic behavior}

The overall magnetic properties of the films were measured using bulk-sensitive superconducting quantum interference device (SQUID) magnetometry and are similar to those reported in Ref.\ \cite{Harrison2015_Gd}. The magnetization vs temperature curve shows paramagnetic behavior and no opening in the hysteresis loop is observed, with saturation occurring near 3~T at 1.8~K. 

\begin{figure}[hb!]      
	\begin{center}
		\includegraphics[width=0.45\textwidth]{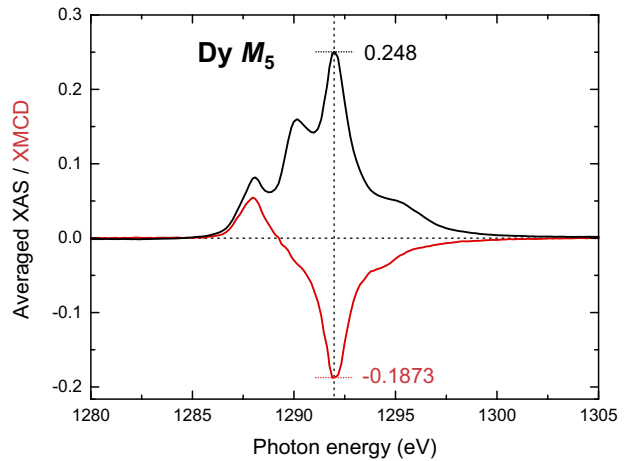}
		\caption{Averaged-XAS and XMCD spectra measured at the Dy $M_5$ edge of the $x=0.31$ sample. The XMCD peak asymmetry is a measure of the effective Dy moment of 3.3~$\mu_\mathrm{B}/\mathrm{atom}$ at 4~K under an applied field of $2\,\mathrm{T}$.}
		\label{fig:Dy-XMCD}
	\end{center}
\end{figure}

\begin{figure}[tb]
	\includegraphics[width=\linewidth]{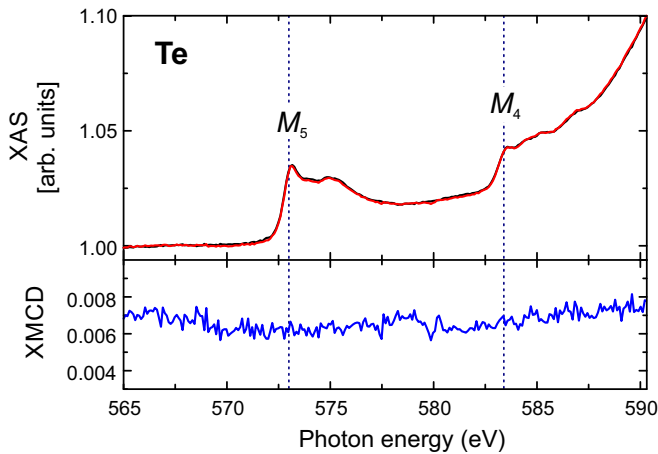}
	\caption{Normalized XAS (top) and XMCD (bottom) intensities for the Te $M_{4,5}$ edges at an applied field of 8~T at 4~K. No spin polarization is detectable on the Te site.}
	\label{fig:M2664_Te_XMCD}
\end{figure}

The temperature-dependent magnetic response was investigated using XMCD at the Dy $M_{4,5}$ edges. There is a clear XMCD signal at 4~K under an applied field of 2~T, see Fig.\ \ref{fig:Dy-XMCD}. 
Determining the asymmetry $A = I_\mathrm{XMCD} / I_\mathrm{SUM} = \left( I_\mathrm{-1} - I_\mathrm{+1} \right) / \left( \frac{2}{3} I_\mathrm{ISO} \right)$
at the $M_5$ peak maximum to be $A_\mathrm{exp} = -0.3776$, and comparing it to the 
theoretical value $A_\mathrm{theory} = -1.217$ calculated for the Dy Hund's rule ground state which has an effective magnetic $4f$ moment $\mu_\mathrm{eff} = 10.65$~$\mu_\mathrm{B}$/atom \cite{FIGUEROA2017}, we obtain $\mu_\mathrm{eff} = 3.3$~$\mu_\mathrm{B}$/atom (31\% of the $A_\mathrm{theory}$ value) at 4~K in a field of 2~T.
XMCD measurements at the Te $M_{4,5}$ edges at the same temperature in an applied field of 8~T showed no measurable response, see Fig.\ \ref{fig:M2664_Te_XMCD}. 

We carried out XMCD measurements at the Dy $M_5$ edge as a function of temperature. The temperature-dependent magnetization curves, $M^2$ vs.\ $H/M$ Arrott plots (not shown) do not follow a linear behavior and no evidence for a ferromagnetic transition was observed, consistent with \cite{FIGUEROA2017}.

\begin{figure*}[tb]
	\includegraphics[width=\linewidth]{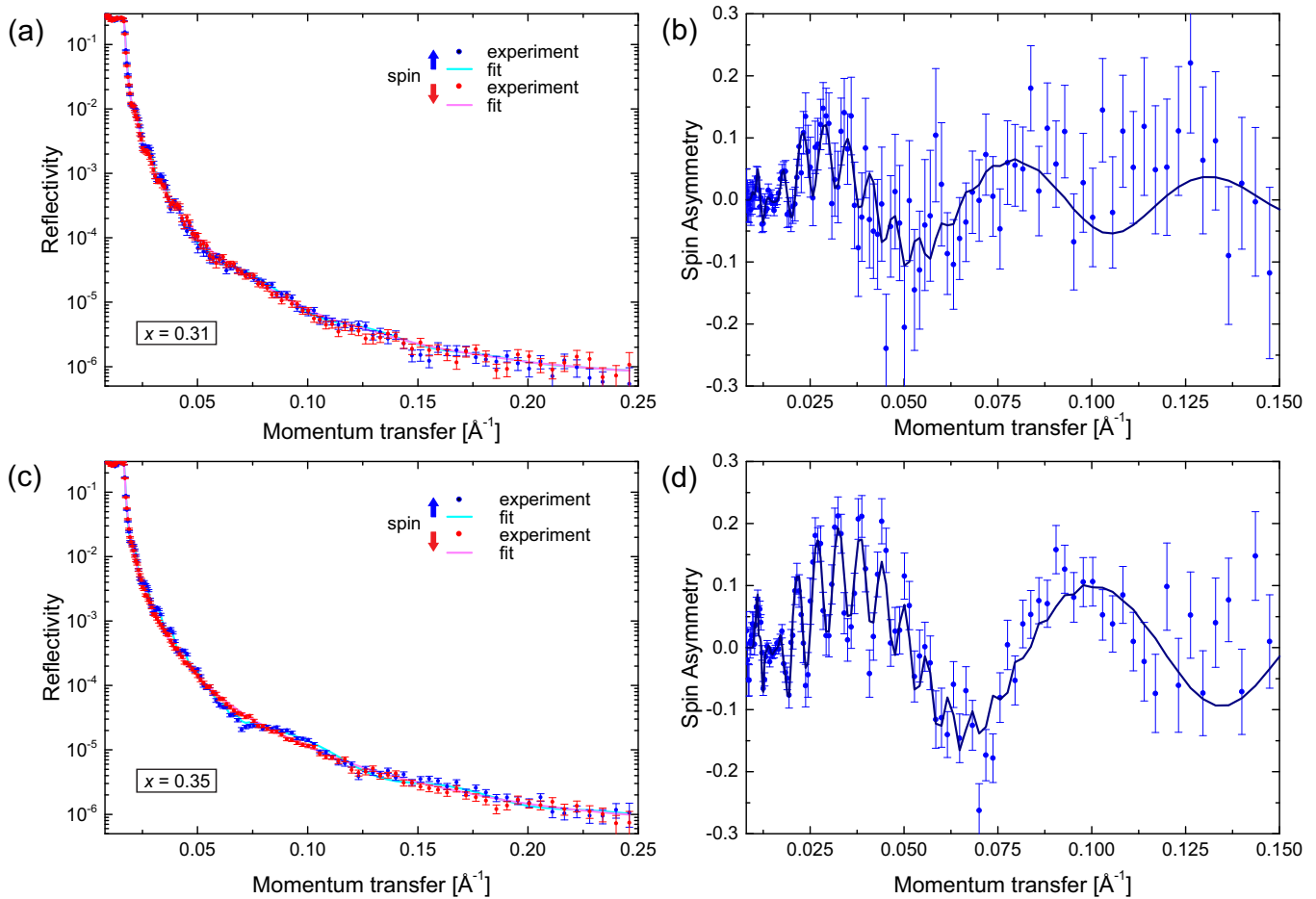}
	\caption{
	(a) PNR data and fit of the $x=0.31$ sample at 5~K under and in-plane applied field of 0.7~T, spin-up blue and spin down red.
	(b) Corresponding spin asymmetry and fit, the normalized difference between the two spin-dependent reflectivities.
	(c) PNR data and fit of the $x = 0.35$ sample at 3~K under and in-plane applied field of 0.7~T, and, (d) corresponding spin asymmetry and fit.}
\label{fig:M2664_PNR}
\end{figure*}

PNR is sensitive to the overall magnetic field $\mathbf{B}$ in the sample, and for a typical paramagnet away from any phase transitions and in moderate applied fields the total internal field remains too weak too be detectable by neutron reflectometry. In contrast, the samples under investigation here show a clear magnetic response at moderate applied fields, 0.7~T, and low temperature, 3~K and 5~K, respectively. Figures \ref{fig:M2664_PNR}(a,c) show the reflectivity curves for the two Dy-doped samples with clear spin splitting in both samples and sizable spin asymmetries, see Figs.~\ref{fig:M2664_PNR}(b,d).
Sizable oscillations are present with maximum amplitudes around 0.15 and 0.2. The effective magnetic moment measured along the quantization axis of the neutron is 1.7~$\mu_\mathrm{B}$ per Dy atom for $x = 0.31$.
For the $x = 0.35$ sample at 3~K the magnetic response is even higher, 2.8~$\mu_\mathrm{B}$ per Dy atom, likely because of the lower measurement temperature. The model fits give an even magnetization throughout the doped layers.
The measurements clearly show that a large effective magnetization, here more than $1/4$ of the full Dy-moment, can be induced even though the TI does not itself order ferromagnetically at these temperatures. 

\begin{figure}[tb]
	\includegraphics[width=7.5cm]{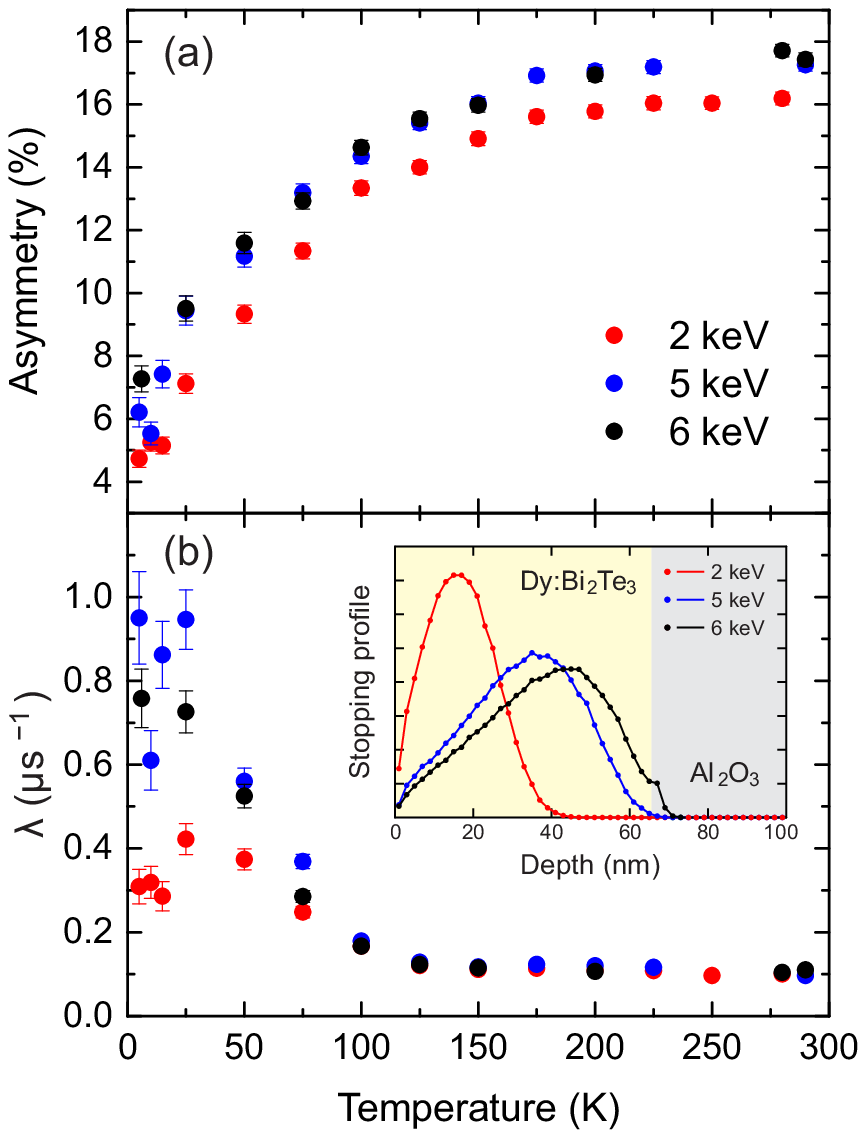}
	\caption{ 
	Temperature dependence of the fitted parameters for TF $\mu^+$SR measurements on samples with $x=0.31$ in an applied field of $B_{0}=5$~mT at several implantation energies.
	(a) Asymmetry amplitude of the oscillatory component of the signal. 
	(b) Relaxation rate $\lambda$. The inset shows the simulated $\mu^+$ stopping profiles.}
	\label{t_dep}
\label{fig:musr_tdep}
\end{figure}

$\mu$SR measurements reveal a complex picture: Transverse field measurements were made on a different $x = 0.31$ film as a function of temperature and implantation energy in a small applied field of $B_{0}=5$~mT. The general behavior is that, (i) oscillations occur at frequencies corresponding to magnetic fields $B$ close to the applied field $B_{0}$;
(ii) the oscillation amplitude remains nearly constant upon cooling from 300~K to 175~K, then it decreases steadily down to 25~K and saturates below this temperature, see Fig.~\ref{t_dep}(a);
(iii) the relaxation rate of the oscillation is seen to remain fairly small until the sample is cooled below $\sim$100~K, at which point it increases until around 25~K, below which it remains high, see Fig.~\ref{t_dep}(b). A second sample with approximately three times lower doping concentration was measured with qualitatively the same behavior. Additional zero field measurements also do not show evidence of long range magnetic order.

The decrease in the oscillation amplitude is consistent with the sample giving rise to a magnetic response, whose volume fraction increases as the temperature decreases. More specifically, the sample appears to contain paramagnetic regions and islands of static magnetism, on the $\mu^+$ ($\mu$s) timescale, which itself might contain some degree of frozen disorder or short range order. Alternatively, this behavior would also be commensurate with fluctuating spins in a paramagnetic environment, which slow down as the temperature decreases. However, supporting longitudinal field measurements (not shown) indicate that the sample contains some regions that give rise to static local magnetic fields. This picture is also consistent with the PNR measurements (see Section \ref{sec:disc}). From similar $\mu$SR investigations \cite{Steinke2018} we know that the residual asymmetry that can be achieved for samples grown in our MBE chamber on Al$_2$O$_3$ substrates and Ni sample plates is of the order of 1-3\,\% for a sample that has more or less fully developed magnetic order. The larger low temperature asymmetry-amplitude in our Dy-doped sample then points towards a fraction of the sample in which no magnetic order develops, i.e. the regions between the islands. A qualitatively similar behavior has been observed for transition metal (V and Cr) doped TIs where the ferromagnetic transition proceeds by the formation of ferromagnetic patches and where it was shown that very slow fluctuations persist well below the $T_\mathrm{C}$ \cite{Krieger2017,Lachman2015,Steinke2018}. However, in those materials, a clear ferromagnetic transition with a well-defined $T_\mathrm{C}$ was observed. This is not the case for Dy-doped samples. 

The residual relaxation rate of the slowly relaxing component, Fig.~\ref{t_dep}(b), increases below 100~K as the amplitude decreases, also consistent with an increase in the static or dynamic disorder. These are both reflected in the relaxation rate.
Interestingly, there is a depth dependence to the residual relaxation rate $\lambda$ with a maximum around 5~keV in the center of the film, as can be seen in Fig.~\ref{t_dep}(b), and as confirmed by a more detailed depth-dependent scan at 5~K (not shown). This implies that the field distribution is not uniform as a function of depth in the film.

\subsection{\label{sec:electron}Electronic properties}

\begin{figure}[t!]
	\includegraphics[width=\linewidth]{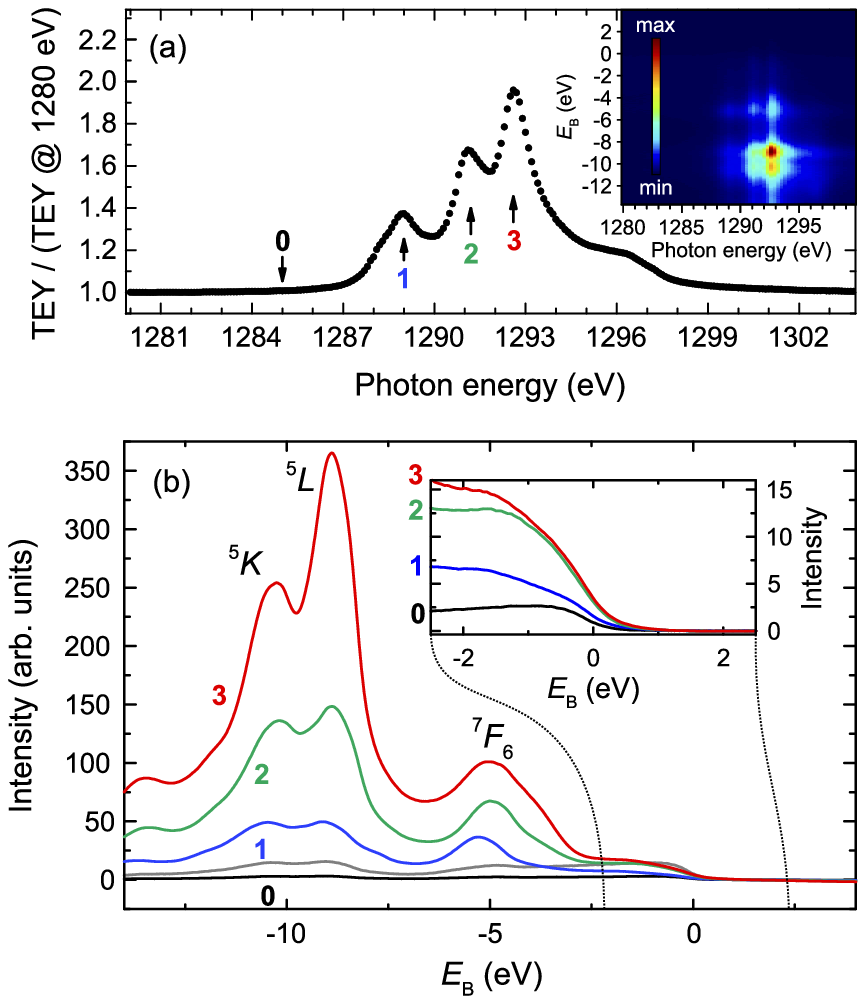}
	\caption{ 
	(a) ResPE in a scan of the photon energy across the Dy M$_5$ resonance (see inset) and corresponding XAS signal (total electron yield).
	(b) ResPE at different photon energies (labeled 0-3) corresponding to the XAS peaks.
	The gray line shows the off-resonance measurement scaled by a factor of 5.
	The term symbols $^7F_6$, $^5L$, and $^5K$ characterize the main multiplet peaks involved in the photoemission from Dy.
	The inset shows a close-up of the ResPE spectra around the Fermi level.
	}
	\label{fig:ARPES}
\end{figure}


The XAS spectrum at the Dy $M_5$ edge (see also Fig. \ref{fig:Dy-XMCD}) is shown in Fig.~\ref{fig:ARPES}(a) together with the corresponding resonant photoemission (ResPE) intensity map as a function of photon and electron binding energies. We restrict our analysis here to angle-integrated PE data because angle-resolved measurements show only a negligible $\mathbf{k}$-dependence, which may be due to the disorder introduced by the doping. Scanning of the photon energy across the Dy $M_5$ absorption edge allows for resonant photoemission, where after an initial x-ray absorption, the core-hole excitation decays emitting an electron. For further details see Appendix A. The XAS and ResPE spectra exhibit a linear dichroism at the peak marked with the label 2 upon changing from $\mathit{s}$- to $\mathit{p}$-polarized incident X-rays (not shown). A similar dichroism in XAS has previously been reported in non-ferromagnetic Dy \cite{Sacchi1991,Sacchi1992}.
Further details of our data are revealed in Fig.\ \ref{fig:ARPES}(b) which presents cuts of the ResPE intensity at photon energies corresponding to the individual peaks of the XAS spectrum shown in Fig.\ \ref{fig:ARPES}(a). 
The peaks shown in Fig.\ \ref{fig:ARPES}(b) at the binding energies of $-5$, $-9$, and $-10.5$\,eV correspond to the multiplet states $^7F_6$, $^5L$, and $^5K$, respectively, which in photoemission are accessible from the $4f^9$ ($^6H_{15/2}$) ground state (see Ref.\ \cite{vdL1993}).
Excitation at the photon energy of a deep core level enhances specific direct photoemission peaks by two orders of magnitude, where the intensities are governed by the dipole matrix elements. The lower energy peak (\#1) in the Dy $M_5$ XAS favors decay into high-spin (septet) final states, whereas the higher energy XAS peaks (\#2 and \#3) favor low-spin (quintet) final states. The structure between $-5$ and $-15$\,eV therefore confirms the local atomic information about the Dy ions.
However, if valence states have an admixture of Dy weight, they will show an  intensity increase at energies near the Fermi level $E_\mathrm{F}$ beyond the multiplet region. We find that there is indeed a small resonant enhancement of the spectral weight at $E_\mathrm{F}$ in the inset of Fig.\ \ref{fig:ARPES}(b). 
The possible presence of Dy excitations close to $E_\mathrm{F}$ can induce charge carriers. We note that the sample under investigation here is very highly doped and includes not only substitutional dopants but also Dy accumulation in the van der Waals gap. We are unable to separate the contributions from these two sites.

\section{\label{sec:disc}Discussion}

Despite the lack of a ferromagnetic transition, our samples display a large internal magnetic field at moderate applied fields and low temperatures; more than $1/4$ of the full Dy magnetization at 3~K and 0.7~T. PNR is a static measurement and so the moments extracted are due to a net increase of the alignment of the Dy-moments with the external field and not due to a slowing of the fluctuations of these moments. The gradual decrease of the asymmetry with temperature in the $\mu^+$SR data reflects the increase of the magnetic volume fraction which we therefore attribute to the formation of magnetically ordered islands embedded in a paramagnetic environment. Below 25~K the magnetic volume fraction remains constant, at least down to liquid He temperatures. The $\mu^+$SR slow relaxation rate probes the field distribution inside the paramagnetic region of the sample. It similarly increases smoothly with decreasing temperature, before reaching a maximum at around 25~K and then remaining high, but with large scatter. This implies an increase in the width of the magnetic field distribution (implying an increase in the magnetic fields) in the paramagnetic region and/or a slowing down of the fluctuation rate of those fields. The magnetic fields in the paramagnetic region can be attributed to the stray fields emanating from the magnetically ordered islands. Again the sizable static magnetization seen in the PNR data is commensurate with an increase in the net magnetic fields but of course a slowing down of the fluctuation rate may also be present. 

We also note that there is a large difference in the effective magnetic moment per Dy-atom in the PNR measurements between 5~K and 3~K, albeit on different samples (see discussion below), whereas the magnetic volume fraction seen in the $\mu^+$SR-asymmetry remains constant below 25~K. This is also consistent with the picture of inhomogeneous islands of some magnetic order, where the Dy-spins order sufficiently below 25~K to fully depolarize the spin of the implanted muons, and of interstitial regions where no magnetic order is present. The muons stopping in the islands no longer contribute to the precession signal. But it is precisely the degree of alignment within (and between) the islands that will give rise to the internal magnetic field that is measured with neutron reflectometry. 

As discussed in detail in Ref.~\cite{Harrison2015_Gd}, the saturation magnetization of Dy-doped samples is strongly doping concentration-dependent, with a strongly decreasing moment for higher doped samples. This implies that the magnetic behavior stems from the substitutional Dy ions rather than the residual accumulations in the van der Waals gap. It is for this reason that we ascribe the majority of the sizable difference in the magnetization measured by PNR to the different acquisition temperatures. 

In contrast to the clear magnetic response on the Dy site, our XMCD measurements could not detect any magnetic signal on the Te site. This is in clear contrast to transition metal doped ferromagnetic TIs where such measurements have shown a measurable polarization on the anion site, confirming a conduction band mediated magnetic response in those systems \cite{PhysRevMaterials.1.064409,Ye2015,Duffy2017} that is absent in the Dy-doped case. Despite this, our ResPE results provide some evidence for possible dopant-induced charge carriers, but these do not seem to contribute to the magnetic ordering as evidenced by the XMCD results. The samples under investigation here are, however, highly doped and lower doping might significantly reduce or perhaps eliminate these additional excitations. As reported in Ref.~\cite{Harrison_SciRep}, we were able to observe a sizable band gap at doping concentrations nearly $3\times$ lower than those investigated here. In addition, the similar ionic radii of Dy$^{3+}$ and Bi$^{3+}$ should also reduce the number of structural defects the dopant introduces. These can also lead to additional charger carriers.\\

\section{\label{sec:concl}Conclusion}

In conclusion, we have shown that the rare earth element Dy is an interesting dopant for 3D TIs. The structure is robust, even for high levels of doping, and no ferromagnetic impurity phases are formed. The Dy-doped TIs are not fully ferromagnetic but instead display complex, short range and inhomogeneous magnetic order. Nevertheless, PNR has clearly shown that a large effective magnetization can be induced. This can be easily exploited in proximity-coupled heterostructures, rather than using external applied fields as used in this study. While the use of Dy as a dopant does not rule out the presence of dopant-induced charge carriers, these do not stabilize the magnetic order. Therefore independent control of these two essential parameters for QAHE applications should be feasible in Dy-doped TI based heterostructures.\\ 

\begin{acknowledgments}
We thank the ISIS neutron source (\url{doi:10.5286/ISIS.E.82353238}) and the Swiss muon source for beamtime, and we acknowledge XMCD beamtime on ID32 at the European Synchrotron Radiation Facility (ESRF) awarded under proposal HC-2718.
T.H.\ acknowledges funding from the John Fell Fund (University of Oxford) and thanks RCaH for their hospitality. L.D.\ acknowledges financial support STFC (UK) and T. L and L. D. from EPSRC (UK). J.A.K., Z.S. and V.N.S.\ acknowledge support from the Swiss National Science Foundation (SNF-Grant No.\,200021\_165910).
\end{acknowledgments}

\section*{Appendix A: Resonant photoemission at the Dy $M_5$ absorption edge}

Scanning the photon energy across the Dy $M_5$ absorption edge allows for resonant photoemission.
For Dy, this second-order process of an initial x-ray absorption and the core-hole excitation decay emitting an electron is given using multi-electronic configurations as $4f^9 + h \nu \rightarrow 3d^9 4f^{10} \rightarrow 4f^8 +\epsilon$, which is in resonance with the direct photoemission process $4f^9 + h \nu \rightarrow 4f^8 + \epsilon$, where $\epsilon$ is a continuum state.
This process enhances the Dy contributions to the total PES spectral weight \cite{Molodtsov1997,Kobayashi2014}.
In the case of rare-earth elements, the strongly localized $f$-electrons are highly correlated, and the spectrum observed by photoemission can be described in the limit of localized atomic multiplet excitations \cite{Arenholz1995,Taguchi2003}.
The resulting spectrum is dominated by the large $4f$-$4f$ Coulomb interactions, which for Dy splits the $4f^8$ final state into a multiplet $\sim$10\,eV wide (see Refs.\ \cite{vdL1993,vdL1999}).

%
\bibliography{DyBiTe_PNR_v17}



\end{document}